# Time-dependent photoionization of azulene:
# Optically induced anistropy on the femtosecond scale

Kevin Raffael, Valérie Blanchet, Béatrice Chatel, Giorgio Turri, Bertrand Girard

Laboratoire Collisions Agrégats Réactivité (UMR 5589, CNRS -Université de Toulouse, UPS),

Institut de Recherches sur les Systèmes Atomiques et Moléculaires Complexes, France

and

Ivan Anton Garcia, Iain Wilkinson, Benjamin J Whitaker

School of Chemistry, University of Leeds, Leeds, LS2 9JT, UK

Corresponding author :

V. Blanchet, fax # : 00 (33) 561 558 317, val@irsamc.ups-tlse.fr

Laboratoire Collisions Agrégats Réactivité (UMR 5589, CNRS -Université de Toulouse, UPS), Institut de Recherches sur les Systèmes Atomiques et Moléculaires Complexes, France

ABSTRACT

We measure the photoionization cross-section of vibrationally excited levels in the $S_2$ state of azulene by femtosecond pump-probe spectroscopy. At the wavelengths studied (349-265 nm in the pump) the transient signals exhibit two distinct and well-defined behaviours: (i) Short-term (on the order of a picosecond) polarization dependent transients and (ii) longer (10 ps – 1 ns) time-scale decays. This letter focuses on the short time transient. In contrast to an earlier study by Diau *et al.*(1) we unambiguously assign the fast initial decay signal to rotational dephasing of the initial alignment created by the pump transition.

Keywords: azulene, rotational coherence, excited state dynamics, femtosecond spectroscopy, doubly excited states

# I. INTRODUCTION

The photochemistry of azulene has been the subject of extensive experimental and theoretical studies due to its second electronically excited singlet state $S_2$ exhibiting a much stronger fluorescence than the lower lying first electronically excited singlet state $S_1$.[2-4] As such, the molecule is well known as the textbook exception to Kasha's rule that "the emitting level of a given multiplicity is the lowest excited level of that multiplicity".[5] Azulene has subsequently become a model compound with which to study intramolecular dynamics and energy transfer processes.[6] Intriguingly, sometime ago Diau *et al.*[1] observed a fast exponential component in the ion time-profile with a 350 fs decay on the top of a long decay component, when exciting the $S_2$ state with 471 meV excess vibrational energy. They suggested that the appearance of the fast 350 fs signal was a direct manifestation of a non-radiative process occurring in $S_2$. However, this is surprising when one considers the ns lifetime time of the $S_2$ state obtained by fluorescence quantum yield experiments[4] and other time-resolved experiments.[2,3,7] This contradiction prompted us to reinvestigate the short-time dynamics in $S_2$ state by time-resolved photoelectron/photoion imaging spectroscopy. These experiments have been performed as a function of the pump wavelength, the relative polarization of the pump and probe pulses and the carrier gas pressure. Our results lead us to conclude that the short decay first observed by Diau *et al.*[1] in fact corresponds to anisotropy in the rotational distribution of the pumped $S_2$ levels.

## II. EXPERIMENTAL DETAILS

We employ pump-probe time-of-flight (TOF) mass-spectrometry to detect either photoions or photoelectrons resulting from the absorption of one pump photon to populate an excited state of the neutral which is then ionized at a controlled time delay by two or more probe photons. As more fully described elsewhere(7) both pump and probe pulses are generated from a 1 kHz-2.5 mJ chirped pulse amplifier (CPA) centered at ~805 nm delivering a fundamental pulse with a Fourier limited Full-Width at Half Maximum (FWHM) of ~60 fs (Amplitude Systems). A Non-collinear Optical Parametric Amplifier (NOPA)(8,9) with subsequent sum frequency mixing or second harmonic generation (SHG) produces the ultraviolet pump pulse. The NOPA output is partly recompressed by a prism pair. The tunability of the NOPA allows the pump central wavelength to be varied from 370 to 265 nm. The typical spectral width of the pump pulse is 250-450 cm$^{-1}$, corresponding to a Fourier limited pulse of 35-60 fs at FWHM in intensity. For all of the experiments reported here the probe pulse is obtained by SHG of the fundamental output of the CPA. The cross-correlation time, recorded by off-resonant multiphoton ionization of nitric oxide, has a typical 120 to 320 fs FWHM depending on the pump wavelength. At the pulse energies employed (<1μJ for the pump and ~15-25μJ for the probe) we are confident that multiphoton ionization takes place in an unperturbed energy scheme.(10)

The focused laser beams intersect, perpendicularly, a skimmed continuous molecular beam containing azulene diluted in nitrogen. This yields photoions and photoelectrons which are subsequently detected, depending on the sign of the applied accelerating voltage, at the output of a 40 cm time-of-flight velocity map imaging spectrometer whose axis is perpendicular to the plane defined by the laser and molecular

beams.(11) The translational speed of the molecular beam was estimated from the length of the azulene ion image and its distance from the centre of the background gas spot due to azulene at rest in the laboratory frame. The distribution of this translational speed corresponds to a temperature of 25 K, and we assume that the rotational temperature is similar. We also performed some experiments in an unseeded beam of pure azulene vapour. Under these expansion conditions we estimate a rotational temperature close to 170 K. We were unable to detect any trace of van der Waals clusters such as azulene-$N_2$ or (azulene)$_n$ in the mass spectrum under either expansion condition.(12,13)

For each pump-probe delay, controlled by a programmable translation stage placed in the pump beam path, an average over ~2000 laser pulses was made to record the transients on the ion signal. Photoelectron spectra were obtained by velocity map imaging over ~ $4 \times 10^5$ laser pulses.(14)

## III. RESULTS AND DISCUSSION

**A-Time-resolved photoionization.**

Figure 1 shows time profiles of the photoion signal of the parent azulene ion, $C_{10}H_8^+$, recorded for different pump wavelengths (265-367 nm) with a 400 nm probe pulse over a 4 ps delay range. All the decay traces were recorded with parallel pump and probe polarisations. The pump excitations span an energy range of over 1 eV from the band origin of $S_2$ ($2A_1$) (Fig.1.c-e) to vibrationally excited levels in $S_4$ ($3A_1$) (Fig.1.a-b).(15) The off-resonance cross-correlation signal (Fig.1 f) gives a FWHM of 320 fs. This cross-correlation time is reduced to 120 fs when using SHG instead of frequency mixing to produce the pump pulse. Whatever the pump excitation, a fast component is evident in all the time profiles (Fig. 1 a-e). It can be fitted by a single exponential decay of 465 fs as illustrated by the solid line in Fig. 1-a. This decay is the same for all pump wavelengths investigated and is in reasonable agreement with the 350 fs decay reported previously by Diau *et al.*(1) Taking into account the difference in electronic character between the $S_2$ and $S_4$ states and the change of vibrational density over the 1 eV energy window scanned here it now seems unlikely that the recurrent fast transient signal can be due to fast internal conversion from $S_2$ or $S_4$ or to dissipative intramolecular vibrational-energy redistribution, as originally suggested by Diau *et al.*(1) (from their ion transient signal obtained at a single excitation energy).

In order to further investigate the physical origin of the photoion time profiles, the transients were recorded with the pump and probe polarizations parallel and perpendicular to one another. Fig. 2 shows a comparison of the time profiles recorded at 335 nm for the two polarization configurations for the azulene cation (top panel) and the fragment $C_8H_6^+$ (lower panel). At this pump energy, the vibrational excitation in $S_2$ is 1093 cm$^{-1}$, slightly lower than the 3800 cm$^{-1}$ investigated by Diau *et al.* (1) The $C_8H_6^+$ fragment ion is produced by dissociative ionization and appears as the intensity of the 400 nm probe is increased to ~25 µJ pulse$^{-1}$. The appearance potential of the $C_8H_6^+$ ion is known to be around 14 eV,(16) and consequently the $C_8H_6^+$ signal must be a product of at least three photon absorption of the probe pulse. As previously demonstrated,(7) power dependency studies on the parent ion support a 1+2' ionization scheme over all the pump wavelength range presented here. The result we wish to draw attention to is that both pump-probe transient signals show a marked and similar difference for the two polarization configurations, with the main change being the apparent suppression of the fast component when the laser polarisations are perpendicular to each other. Experiments were also performed for a relative polarisation angle of 54.7°, the magic angle, which cancels any anisotropic dependency induced by the pump on the pump-probe signal. These confirmed the polarisation dependence of the fast signal.

Such polarization dependent transients are expected for an optically induced anisotropy investigated by time-resolved polarisation sensitive spectroscopy.(17-22) The effect can be described classically as the dephasing of an ensemble of molecular rotors (whose initial population is defined by the statistical distribution at the rotational temperature of the molecular beam, the Höln-London factors and the selection rules $\Delta J= 0,\pm 1$). Photons polarised with respect to the laboratory frame excite only those molecules in the sample (in our case a "rotationally warm" molecular beam) whose transition dipoles are favourably aligned with respect to the laboratory frame. This results in a macroscopic polarization of the sample. Because the population of initial states is determined by Boltzmann energy partitioning, the excited ensemble is made up of molecules with a range of rotational states whose classical rotational frequencies are not the same and consequently the macroscopic polarization of the ensemble rapidly becomes misaligned. The initial coherence between the rotational states in the ensemble is then lost.

Rotational coherence spectroscopy (RCS) may be performed in a variety of ways, for example, by fluorescence,(17-19) multiphoton ionization or photoelectron imaging.(22) However, it is most easily observed in linear and symmetric top molecules because in asymmetric rotors the revival structures are more complicated and the signals consequently weaker. Despite this a number of observations of rotational revivals in asymmetric molecules have been made using emission and ionization techniques.(23) Enhanced sensitivity of RCS in asymmetric rotors can, also, be achieved by using degenerate four wave mixing as a detection scheme.(24-26) Azulene is an asymmetric rotor ($I_x$=2841.951839(193), $I_y$=1254.843135(66), $I_z$=870.712723(64) MHz)(27) and although we have not been able to observe rotational revival structures in the photoion/photoelectron decay traces we can infer the contribution of the initially induced orientational coherence from the short time behaviour of the pump-probe decay traces.

By rotating the polarization of the probe pulse relative to the pump pulse, the anisotropy of the ensemble can be investigated by polarization sensitive photoionization. Both the $S_2$ and $S_4$ states of azulene have $A_1$ electronic symmetry, and consequently both states have an electric dipole transition moment with the ground electronic state aligned along the *z* axis (long axis in the molecular plane, see Figure 1). The fast relaxation from $S_4$ to $S_2$ within a 120 fs timescale has been discussed elsewhere,(7) and would not perturb this initial alignment. The polarization sensitivity to ionization by the probe pulse is therefore expected to be the same regardless of whether the $S_4$ or the $S_2$ state is populated. The anisotropy function $r(\tau)$ is defined as :

$$r(\tau) = \frac{I_\parallel(\tau) - I_\perp(\tau)}{I_\parallel(\tau) + 2I_\perp(\tau)} \tag{1}$$

where $I_\parallel(\tau)$ and $I_\perp(\tau)$ refer to the photoionization intensities at a time delay $\tau$ for parallel and perpendicular polarization configurations respectively. Since the rotational dephasing time of the excited ensemble depends on the width of the state distribution in the initial rotational wavepacket, $r(\tau)$ relaxes faster with increasing molecular beam rotational temperature. This is clearly illustrated in Fig. 3 for a pump pulse irradiating at 335 nm. The minimum of the anisotropy function for a rigid asymmetric top molecule due to free molecular rotation occurs around $\tau_{min} = \sqrt{I_y / kT_{rot}}$.(28) From the observed values of $\tau_{min}$=1.7 ps and $\tau_{min}$=4.5 ps, the rotational temperatures are calculated to be around 25 K for the seeded beam and 170 K for a pure azulene beam, in very good agreement with our time of flight measurements of the translational temperature of the molecular beam.

The time-dependent ionization of electronically excited azulene studied as a function of the polarization and rotational temperature demonstrates that the short transient observed corresponds to optically induced anisotropy rather than electronic relaxation via internal conversion. Nevertheless, the question arises as to exactly how this sensitivity via the probe photoionization is achieved. Indeed, we record essentially identical pump-probe transients regardless of the excitation energy over a wide spectral window that spans at least two electronic states ($S_2$ and $S_4$). We have previously demonstrated (7) (i) that the $S_4$ state decays very rapidly towards the $S_2$ state within less than the pump pulse duration (120 fs), and (ii) that whatever the pump wavelength the photoionization is driven by a sequential two photon transition from the $S_2$ state. This explains why the same behaviour and contrast is observed for data recorded above and below the $S_4$ threshold. The scenario is summarized in Fig. 4 for the case of a direct $S_2$ excitation. This two photon transition involves a resonance with an intermediate doubly excited state that decays within the duration of the probe pulse to a set of Rydberg states that are subsequently ionized by the second probe photon. This doubly excited state has been detected by magnetic circular dichroism on different derivatives of azulene around 4.95 eV.(29) The photoelectron images show the fingerprint of the Rydberg states,(30,31) which is manifest by the observation that the photoelectron spectrum is invariant to the pump wavelength.(7) The characteristic spectrum is depicted on the right hand side of Fig. 4. The invariance to the pump wavelength arises because the vibrational excess energy in the Rydberg state is conserved in the cation from Franck-Condon arguments. The overall transition moment from $S_2$ to the doubly excited state must lie along the $z$ axis in order to explain the larger ionization probability for parallel polarized pump and probe fields.

To summarize, at all the wavelengths studied the transient photoionization signals exhibit two distinct and well-defined behaviours: (i) Short-term (on the order of a picosecond) polarization dependent transients which are due to optically induced anisotropy where the sensitivity is achieved by photoionizing via a doubly excited state as sketched in Fig. 4 (ii) longer (10 ps – 1 ns) time-scale decays corresponding to the internal conversion of $S_2$ to $S_0$. The latter can be described by a statistical model.(7)

## IV. Conclusions

We have investigated the time-profiles of pump-probe transients in azulene in which the pump photon excites the electronically excited singlet states $S_2$ to $S_4$ and the subsequent dynamics are probed via photoionization. On the contrary to what has been suggested previously,(1) the short term decay behaviour that is observed on the timescale of ~400 fs when the polarizations of the pump and probe laser fields are parallel is associated with an optically induced anisotropy of the molecular axes in the $S_2$ state. This is observed via two-photon ionization that involves a doubly excited state relaxing onto a set of Rydberg states.(7)

**Acknowledgments**

This work was supported by CNRS, le Ministère de la Recherche, Région Midi-Pyrénées through "Plan état-Région Spectroscopies Optiques Ultimes", ANR COCOMOUV and the British Council Alliance programme. BJW thanks the CNRS and the UPS for his invited positions in the LCAR. KR thanks the CNRS for his postdoc fellowship. GT thanks the European network COCOMO for his postdoc fellowship. IW is grateful to the UK's EPSRC for a research studentship.

FIGURE CAPTION
**Figure 1:**

Pump-probe photoionization transients of $C_{10}H_8$ for different pump wavelengths and a probe pulse centered at 400 nm. The pump and probe polarizations are parallel. Signals a-b correspond to excitation into the $S_4$ electronically excited state, whilst signals c-e involve the $S_2$ state. The cross-correlation time measured off-resonance, trace f, is 320±14 fs. These cross-correlation fit is reported in dot lines in each panel. All the transients exhibit two exponentially decaying components. The faster of the two can be fitted by an exponential decay with a time constant of 465 fs. The convolution of the two decays with the cross-correlation signal is shown as the solid line in trace a.

**Figure 2:**

Comparison of the time-resolved photoionization signals for parallel and perpendicular configurations of the laser polarisations. The pump and probe are centred at 335 nm and 400 nm respectively. The transients are recorded on both the parent ion, $C_{10}H_8^+$, (top panel) and the main fragment ion, $C_8H_6^+$, (bottom panel).

**Figure 3:**

Effect of rotational temperature on the anisotropy function *r(t)*. The rotational temperature is estimated from the minimum of the *r(t)* function: $\tau_r = \sqrt{\dfrac{I_y}{kT_{rot}}}$. This leads rotational temperature around 25 K for the 160 Torr $N_2$ beam seeded with azulene and 170 K for the pure azulene jet.

**Figure 4:**

Pump-probe ionization scheme involving a doubly excited (**) state. The one photon pump transition is indicated by the thick black arrow. The transient photoelectron and the azulene ion signals are produced by a subsequent two-photon probe transition, indicated by the two thin arrows. The first probe photon is postulated to be degenerate with a doubly excited state (most probably vibrationally highly excited) which rapidly decays, within the duration of the probe pulse, to high lying vibrational levels of a set of Rydberg states converging onto the ground state of the cation, $D_0$, giving the photoelectron spectrum shown on the right hand side of the figure. See ref. (7) for full details.

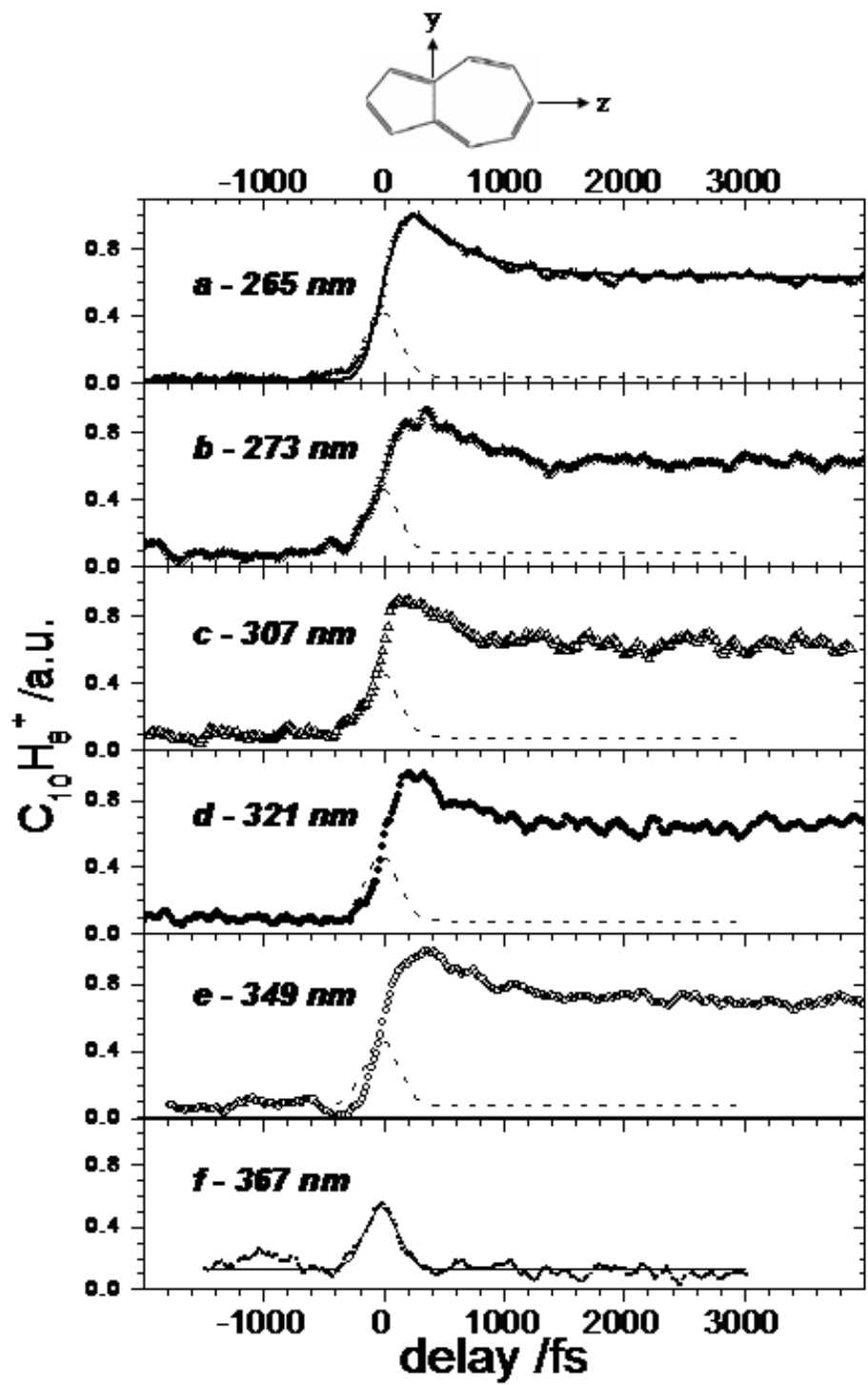

**Figure 1**

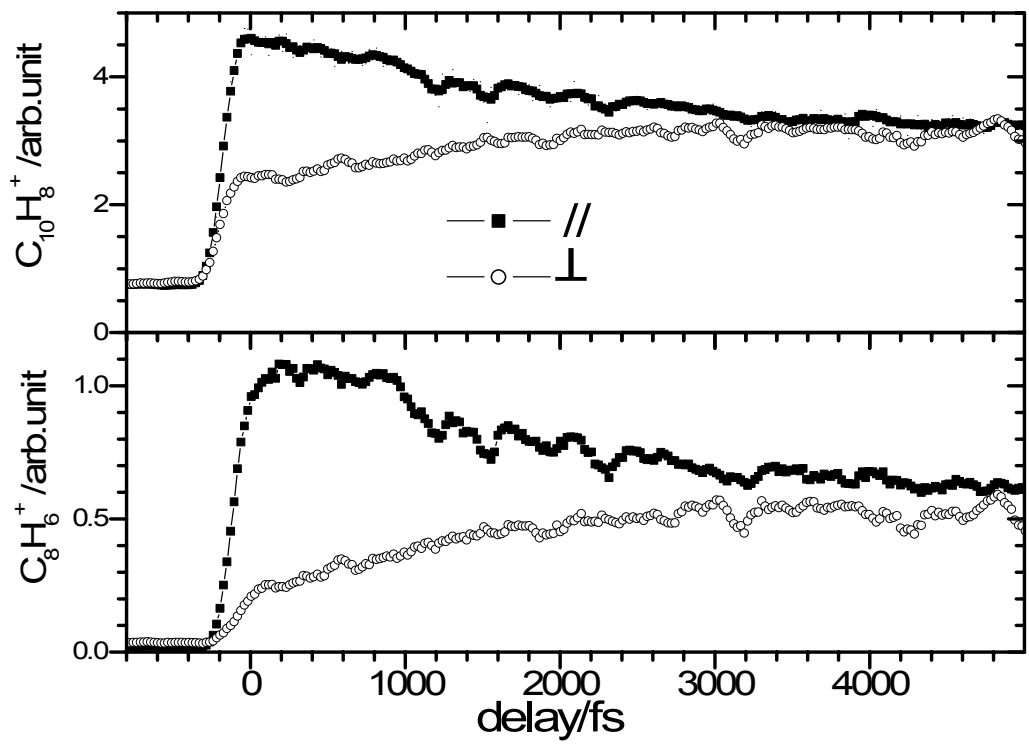

**Figure 2**

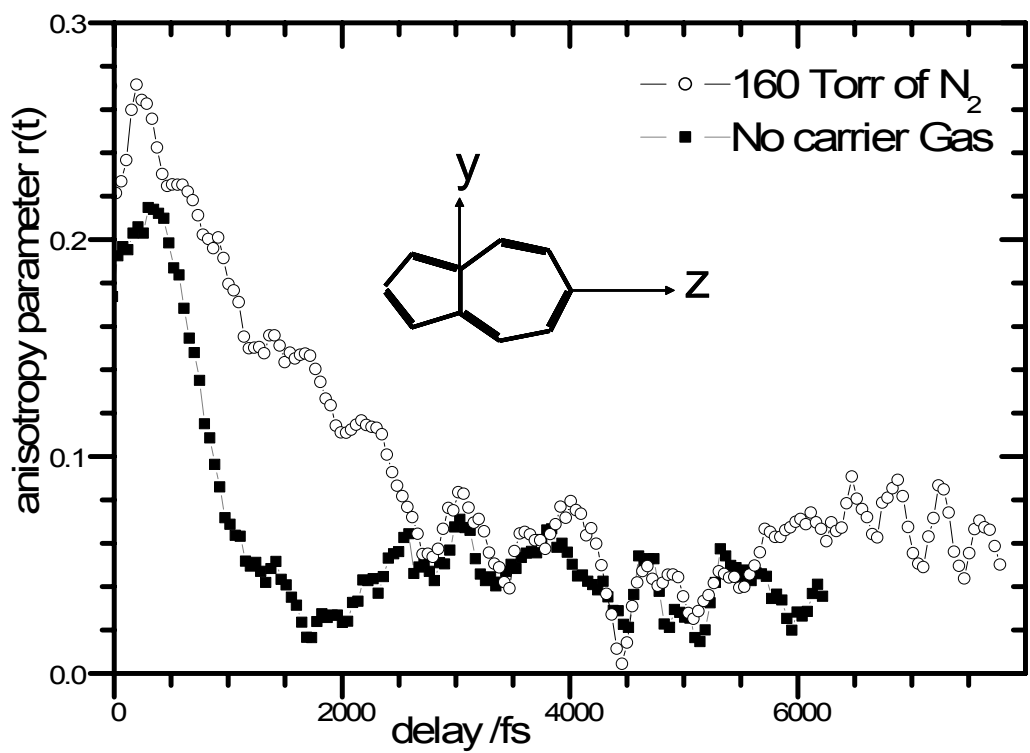

**Figure 3**

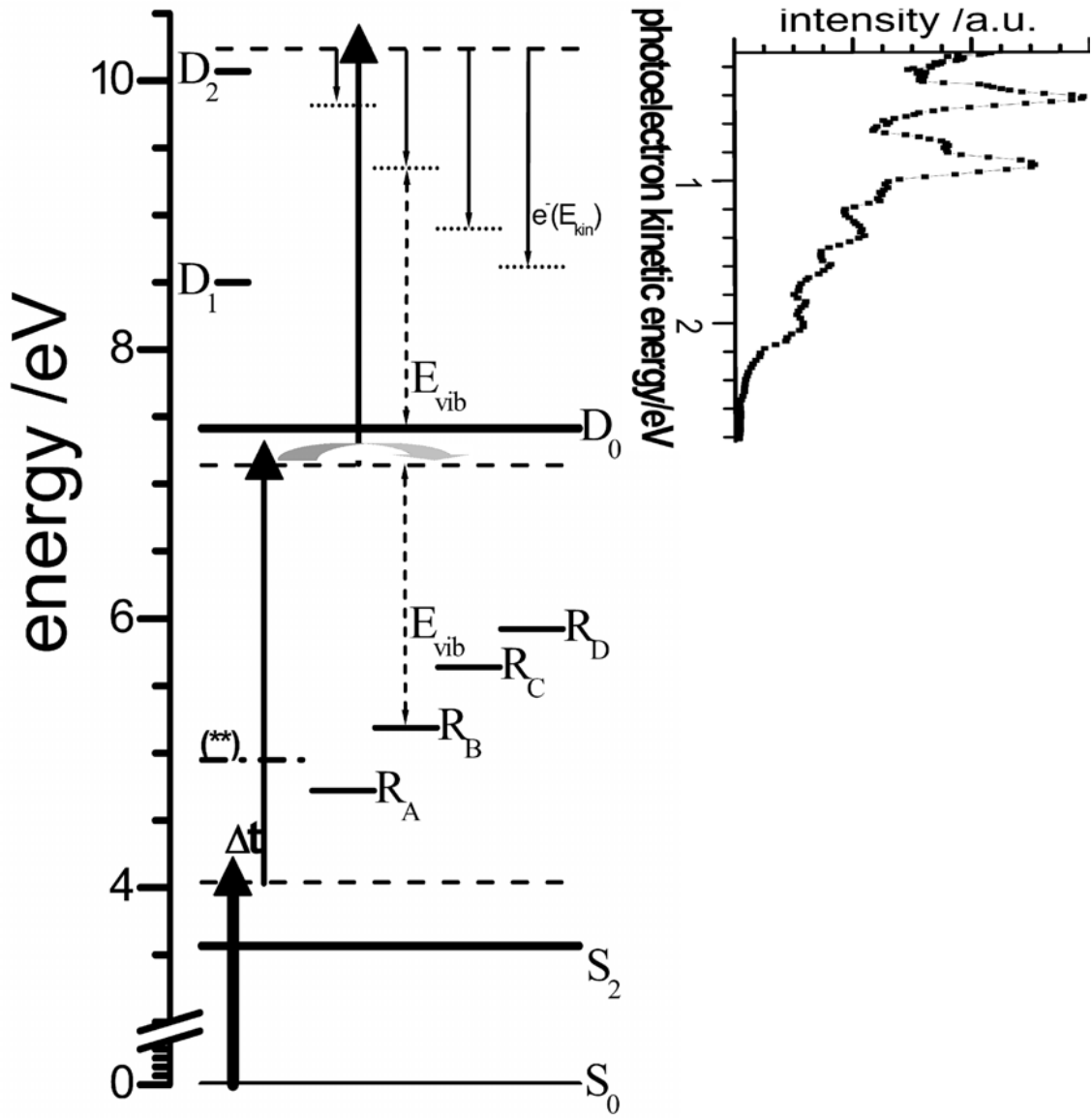

**Figure 4**